# The Very Local Universe in X-rays

A White Paper submitted to the Decadal Survey Committee

**Authors:** A. Ptak[1], E. Feigelson[2], Y.-H. Chu[3], K. Kuntz[1], A. Zezas[4], S. Snowden[5], D. de Martino[6], G. Trinchieri[7], G. Fabbiano[4], W. Forman[4], G. Tagliaferri[7], R. Giacconi[1], S. Murray[4], S. Allen[8], M. Bautz[9], S. Borgani[10], N. Brandt[2], S. Campana[7], M. Donahue[11], K. Flannagan[12], R. Gilli[7], C. Jones[4], N. Miller[1], G. Pareschi[7], P. Rosati[13], D. Schneider[2], P. Tozzi[10], A. Vikhlinin[4]

1. Johns Hopkins University
2. Pennsylvania State University
3. University of Illinois at Urbana-Champaign
4. Harvard-Smithsonian Center for Astrophysics
5. NASA Goddard Space Flight Center
6. INAF–Osservatorio Astronomico di Capodimonte
7. INAF-Osservatorio Astronomico di Brera
8. Stanford University
9. Massachusetts Institute of Technology
10. INAF-Osservatorio Astronomico di Trieste, Trieste, Italy
11. Michigan State University
12. Space Telescope Science Institute
13. European Southern Observatory (ESO)

## *Science Frontier Panels*
**Primary Panel:** Galactic Neighborhood (GAN)
**Secondary Panel:** Galaxies across Cosmic Time (GCT)

*Project emphasized*: **The Wide-Field X-Ray Telescope (WFXT)**
http://wfxt.pha.jhu.edu

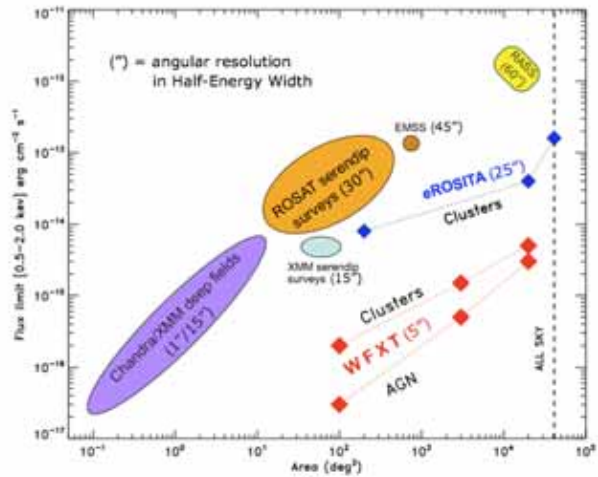

## Executive Summary

While the exceptional sensitivity of Chandra and XMM-Newton has resulted in revolutionary studies of the Galactic neighborhood in the "soft" (<10 keV) X-ray band, there are many open questions. These questions include:
- Does the Local Hot Bubble really exist?
- What is the recent (t <~ $10^7$ years) star-formation rate in the solar neighborhood and in molecular clouds?
- What are the properties of accreting sources and the hot inter-stellar medium (ISM) in Galactic star-forming regions, the Magellanic Clouds and Local Group galaxies?
- What is the (true) correlation between X-ray emission and stellar mass and star-formation rate?
- What is the luminosity function for normal/starburst galaxies?

The common thread between these topics is stellar evolution, from star formation through to the final stage of compact objects, and its "feedback" on galaxy structure and evolution. Chandra and XMM-Newton are not dedicated survey instruments and have relatively small fields-of-view (FOV) and high particle backgrounds. These factors have limited their ability to observe very large (> 10 deg.) objects, low surface brightness objects, and to amass the large numbers of detections of sources that are required to draw firm statistically-robust conclusions. We discuss how very large-angle (> 100 $deg^2$) surveys in the X-ray band will shed light onto these important issues[1]. A deep understanding of the local universe is clearly critical for the proper interpretation of the z>0.1 universe. High-quality very wide-angle X-ray survey data would be revolutionary in their own right for population studies (such as luminosity functions and environment analyses) and would be an excellent complement to existing and planned ground-based wide-area surveys such as SDSS, Pan-Starrs and LSST and supply exotic targets for IXO and Gen-X. The surveys discussed here would be *several orders of magnitude more sensitive* than any performed by other previous or currently planned soft X-ray survey missions.

## Galactic X-ray Emission

Starburst galaxies drive winds and super winds that place large amounts of enriched material into the IGM. The same processes exist in normal galaxies like the Milky Way driving material into the halo, delaying enrichment of the disk, and regulating star-formation, either by bulk motion or turbulence. Star-formation and its tracer, the hot ISM, plays a crucial role in galactic evolution (Veilleux et al. 2005, ARAA, 43, 769). However, our understanding of the hot ISM in the Milky Way, its origins, its spatial distribution, how it gets from its birth places to the places we see it, and just how much of it there is, has not changed since the ROSAT era.

The diffuse Galactic components identified by ROSAT were the halo, the bulge, a disk component, and, possibly, an extremely soft local component called the Local Hot Bubble (LHB). The spatial distribution (excluding remnants or identifiable features such as superbubbles) remains poorly understood. Given that the emission is diffuse with scales of degrees, distance and mass limits can only be determined using shadows due to cool absorb-ing clouds with known distances (Kuntz et al. 1997, ApJ, 484, 245). These shadows also allow spectral isolation of the emission components; the difference between the on- and off-cloud spectra allows one to determine the spectra of both the foreground and background emission. With a sufficiently large catalog of shadows distributed across the high-Galactic-latitude sky (e.g. Snowden et al. 2000, ApJS, 128, 171) one could reconstruct the thermal structure of the Galactic halo as a function of location and significantly constrain its nature and origin (e.g. hydrostatic equilibrium vs. galactic fountain). Determining the spectrum of the local emission is complicated by solar wind charge exchange emission (SWCX), which varies on the same time scales as the solar wind (minutes to days, see e.g., Cravens et al. 2001, JGR, 106, 24883). Spectral isolation of different emission components is thus not possible with *separate* on- and off-cloud measurements as the local spectrum often changes between observations (Snowden et al. 2004, ApJ, 610, 1182). Instead, on- and off- cloud measurements of clouds *must* be done with the same observation, which requires a large FOV and a large grasp.

---

[1] Specifically we address surveys by the Wide Field X-ray Telescope (WFXT; Murray et al. 2008, SPIE, 7011, 46)

The Galactic X-ray Ridge emission (Figure 1), concentrated towards the Galactic center, has a metal-rich X-ray spectrum (at energies above those affected by SWCX) as well as a hard power-law continuum (10-40 keV). Thermal plasmas are a poor explanation for the bulk of the emission, as they could not be gravitationally bound. Possible mechanisms include quiescent X-ray binaries, cataclysmic variables (CVs), coronally active stars, magnetically confined plasmas, inverse Compton scattering from cosmic ray electrons, and non-thermal emission lines from low-energy cosmic ray baryons (Valinia et al 2000, ApJ 543, 733, Ebisawa et al 2001, Science, 293, 1633, Revnivtsev et al 2006, A&A, 452, 169)[2]. The Galactic Ridge would naturally be observed as a by-product of a wide-area Galactic X-ray survey, which would simultaneously resolve faint point sources, measure their composite spectrum, and measure the spectrum of the remaining emission. In addition, scans of several degrees around the Galactic Center and shadowing observations in Baade's Window should reveal the Galactic wind expected from the nucleus (Muno et al 2004, ApJ 613, 326).

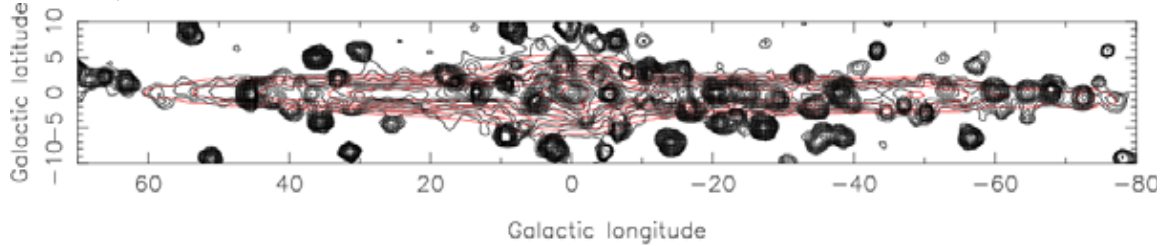

**Figure 1**: Galactic X-ray Ridge. The distribution of X-ray emission near the Galactic center based on 29 Ms exposure with the RXTE satellite. (red: COBE/DIRBE near-infrared map (3.5 $\mu$m) convolved with RXTE response, black: x-ray; Revnivtsev et al. 2006, A&A, 452, 169).

The Gould Belt is a 700 pc diameter ring of star formation regions in which the sun is off-center. Not only does the Gould Belt dominate the local current star formation, but some of its stars are responsible for the Loop I superbubble and may have been responsible for the (putative) LHB. The Gould Belt has been used as a possible instance of self-propagating star formation. Its history, therefore, holds the key to understanding not only the local star-formation history but also the mechanisms governing large-scale star-formation. Radio and infrared surveys trace only the current (< 5 Myr) star formation; X-ray surveys trace the 10-30 Myr population (nearly impossible to distinguish from the field population in any other band) as these stars have X-ray luminosities 100 times higher than field stars (Preibish & Feigelson 2005, ApJS, 160, 390). Given the non-uniformity of Gould's Belt seen in the ROSAT map (Figure 2), the star-formation history likely varies spatially as well, and sensitive wide-area X-ray scans will give a more complete picture the local star formation history.

---

[2] Recent work (Revnivtsev et al. 2009, Nature, submitted) has found that >80% of the Ridge emission has been resolved by Chandra in sources that are likely to be accreting white dwarfs and coronally-active stars.

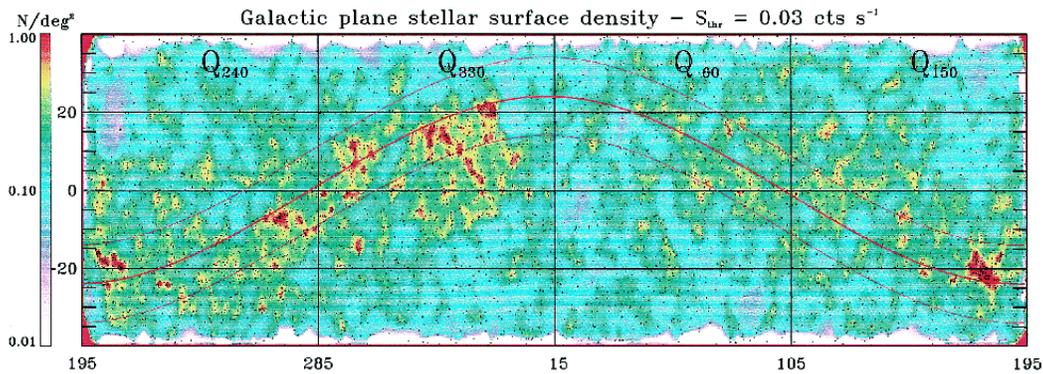
**Figure 2**: RASS image of the Gould Belt

Taking advantage of the elevated flaring of lower-mass pre-main sequence stars and the shocked winds of OB stars, X-ray studies of young stars associated with molecular clouds have been very successful (Feigelson et al. 2007, astro-ph/0602603). Several million seconds of Chandra and XMM-Newton time have been devoted to studies of young stellar populations, including seven Large Projects, which will produce a catalog of >20,000 recently formed stars. These samples enable a host of astrophysical studies concerning the origin of OB stars, the dynamics of stellar cluster formation, triggered star formation, and the survivability of protoplanetary disks in rich clusters. *X-ray surveys are often the only viable way to identify young stars making up the Initial Mass Function*, as optical photometric surveys are often overwhelmed by unrelated old Galactic field stars and infrared photometric surveys are biased towards stars hosting protoplanetary disks.

Existing X-ray surveys are restricted in practice to relatively small fields, $<\sim 1$ deg$^2$ for Chandra and $<\sim 10$ deg$^2$ for XMM-Newton. Molecular cloud complexes are often more extended than this [Reipurth08]: the nearby Ophiuchus complex at $d\sim 150$ pc covers $\sim 30$ deg$^2$ and several star-forming complexes at d < 1 kpc subtend of order $10^2$ deg$^2$ . The rich clusters of the Sco-Cen Association, the nearest OB association whose molecular gas is nearly all dissipated, subtend ~2000 deg$^2$ and its outlying groups are scattered over most of the southern sky.  More distant starburst regions, such as the W3-W4-W5 and Carina complexes, each subtend 10–20 deg$^2$ (Figure 3).

A very wide X-ray survey, supplemented by deeper scans of selected star forming regions,

will give large and unbiased samples of young stellar populations distributed throughout giant molecular clouds. A sensitivity of 1 x $10^{-15}$ erg $s^{-1}$ $cm^{-2}$ in the 2–10 keV band corresponds to $L_X \sim$ 1 x $10^{29}$ erg $s^{-1}$ at $d$=1 kpc, which is sufficient to capture ~40% of the stellar Initial Mass Function at $d$=1 kpc. These samples will elucidate the rate of distributed star formation, the drifting and ejection of young stars from their natal clouds, and recent past episodes of star formation.

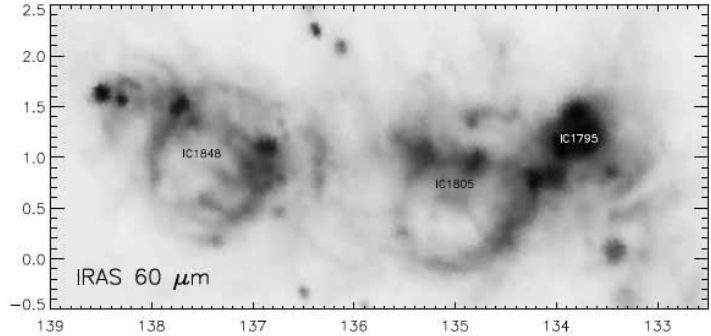

**Figure 3**: Global map of the W3-W4-W5 complex at 2 kpc [Megeath08].

There are large populations of accreting sources that occupy the Galactic plane that can only be exhaustively studied with very wide surveys. These include high-mass X-ray binaries, low-mass X-ray binaries and CVs. Surveys such as these are necessary to detect these sources in sufficient numbers to derive high-quality luminosity functions, particularly for high-mass X-ray binaries since the unique identification of the optical counterpart will not be difficult. The nature of the new class of γ Cas analogues is a still a debated but timely issue as they could be the Be/WD binary population predicted to outnumber the Be/NS systems (Lopes de Oliveira et al. 2006, A&A 454, 265; Motch 2008, AN 329,166). A deep survey will also discover the still hidden population of short period AM CVn binaries, expected to be strong sources of gravitational radiation (Nelemans et al. 2004, MNRAS 349, 181). They represent a still debated candidate channel for Type Ia SNe (Solheim & Yungelson 2005, ASPC 334, 387). For low-mass X-ray binaries these surveys will provide a legacy dataset for telescopes with higher spatial resolution (most notably Gen-X). A sensitive wide area soft X-ray survey is also ideal for identifying isolated neutron stars based on their (almost) perfect blackbody spectra. So far there are only seven such sources known (Haberl F., 2007, A&SS, 308, 171), mainly due to their very low space density and the difficulty in identifying them. Their populations are important for constraining neutron-star formation rates (e.g. Keane & Kraemer, 2008, MNRAS, 391, 2009), while determination of their mass-to-radius ratio (from their thermal emission) is important for constraining their equation of state.

### The Magellanic Clouds
The Large and Small Magellanic Clouds (LMC & SMC) provide an excellent laboratory to study X-ray sources. Their proximity (50 and 60 kpc), nearly face-on orientation, and minimal line-of-sight obscuration make it possible to inventory both X-ray point sources and diffuse emission throughout an entire galaxy, and further associate them with the underlying stellar population. Sources within each galaxy can be inter-compared and their physical parameters can be determined without distance ambiguity.

Figure 4 displays an optical image and a ROSAT PSPC mosaic of the LMC, showing the warm ($10^4$ K) and hot ($10^6$ K) ionized interstellar gas, respectively. Diffuse X-ray emission is detected in SNRs, superbubbles, supergiant shells, and active star forming regions, where interstellar gas is shock-heated by fast stellar winds and supernova explosions. Diffuse X-ray emission is also detected in the field unbounded by obvious interstellar structures, where the heating mechanism is not very clear. ROSAT PSPC mosaics of the MCs, currently the best X-ray images with complete coverage, are severely limited because more than ½ of the surface area was observed only during the ROSAT All Sky Survey with exposures shorter than 1-2 ks. A sensitive high-resolution (5-10") X-ray survey of the MCs will allow us to establish a complete catalog of SNRs and use it to determine supernova rates, study SNR shock heating in isolated environment or in superbubbles and supergiant shells, and map the distribution and determine the physical properties of hot ionized gas throughout an entire galaxy. For a comprehensive investigation of energy feedback, X-ray observations of the hot ionized interstellar gas should be analyzed in conjunction with optical observations of the warm ionized gas and underlying stellar population and radio observations of the neutral atomic and molecular gas (e.g. Cooper et al. 2004). The energy feedback problem is complicated by the discovery of nonthermal X-ray emission in superbubbles, e.g., 30 Dor C (Bamba et al. 2004). To search for diffuse nonthermal X-ray emission, X-ray detectors need to have high sensitivity and low background at energies of 2-5 keV.

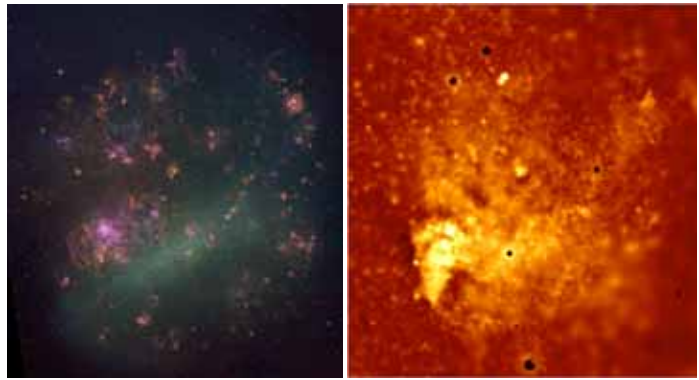

**Figure 4.** Magellanic Emission Line Survey image and ROSAT PSPC mosaic of the LMC.

Sensitive high-resolution X-ray surveys of the MCs can detect point sources with a completeness limit of $10^{32}$-$10^{33}$ ergs/s. The LSST observations of the MCs can be used to identify optical counterparts of the X-ray point sources, not only the HMXBs but also the LMXBs. LSST observations will also reveal the stellar population and star formation history of the MCs. Complete samples of X-ray binaries with known companions and underlying stellar population will allow us to fully test X-ray population synthesis models (Belczynski et al. 2008, ApJS, 174, 223).

The MCs have a trove of X-ray binary pulsars (e.g. Coe et al. 2008, arxiv:0809.2665). The intense monitoring and large field of view of the Rossi X-ray telescope enabled the discovery of many of these pulsars. On the other hand, Chandra and XMM-Newton allowed the detection of X-ray binary pulsars at lower luminosities, but their small FOV limited these studies to specific regions of these galaxies. Most importantly the current studies of extragalactic pulsars are mainly phenomenological and limited to the identification of new pulsars and measurements of their periods. Any further understanding of the evolution of X-ray binaries in low metallicity environments and the physics of accretion onto highly magnetized objects require measurement of their orbital parameters, spin evolution, and studies of variations of the accretion flow during their orbit. Wide-area X-ray surveys will greatly benefit these studies by allowing monitoring of a large fraction of the MCs area. This way we will be able to:

- Constrain the total number of X-ray binary pulsars
- Study their long-term spectral variability which is important for constraining accretion physics in the presence of strong magnetic fields (e.g. the "propeller effect", Shtykovskiy & Gilfanov 2005, A&A, 431, 597)
- Study the evolution of quiescent X-ray pulsars
- Constrain the orbital periods of X-ray binary systems and the long-term spin evolution of the X-ray binary pulsars (constraining their spin-down).

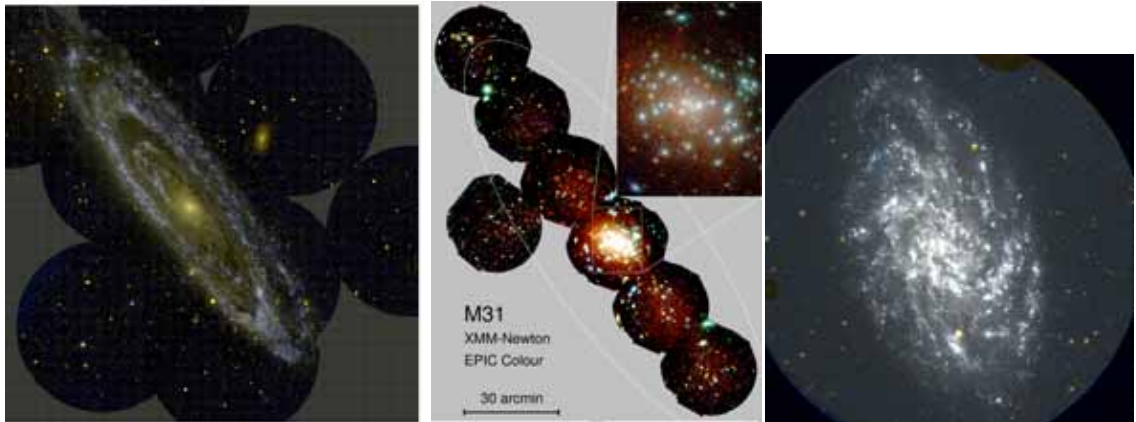

**Figure 5**: M31 observed by Galex (left) and XMM-Newton (middle) and M33 observed by Galex (right; Thilker et al. 2005, ApJ, 619, L67). The XMM-Newton image only shows a subset of the current observations of M31 however the smaller FOV of XMM-Newton requires a much larger number of pointings compared with Galex, which has a FOV similar to WFXT.

### Local Group Galaxies

Current attempts to monitor the X-ray binary populations of M31 with Chandra and XMM-Newton focus only on specific regions of the galaxy, which provides only a partial view of its X-ray source population. Observations of a larger fraction of the galaxy would allow us to detect a much larger number of sources at each epoch and therefore begin to constrain the duty cycles and outburst timescales of X-ray binaries in M31. These are critical (and unknown) parameters for understanding the physics of accretion (the timescales of disk instabilities are intrinsically linked to the duty cycles of X-ray binaries) and evolution of X-ray binaries. A few ks exposure with a survey telescope such as WFXT can reach $10^{35}$ erg s$^{-1}$ at 1Mpc (i.e., to the outskirts of the Local group), which can capture all X-ray binaries in outburst. Similar campaigns in the case of M33 will give a picture of the variability patterns of the younger X-ray binary populations. More generally these types of observations will form a more complete census of transient sources in general such as white-dwarf binary SSS, which may be alternative channel of SN Ia progenitors (Di Stefano et al. 2004,

ApJ 210, 247). Most SSS in M31 are identified with optical novae, with an SSS state duration of order months to years, may imply that the globular cluster optical novae rate may be drastically underestimated (Henze et al., 2009, A&A, arXiv0811.0718). These observations will provide a legacy dataset for future high spatial resolution X-ray telescopes that we will use to identify their counterparts and possibly measure their mass function in order to get a complete picture of the X-ray binary populations in our Local Group.

**Nearby Galaxies**

It has been known since the 1980s that the X-ray emission of spiral and starburst galaxies is proportional to both the current star-formation rate and the total stellar mass (Fabbiano 1989, ARAA, 27, 87). However our knowledge of the precise shape and normalization of the X-ray/SFR and X-ray/mass relationships is uncertain since it is based on a relatively small number of galaxies in mostly pointed observations, and therefore is potentially biased by X-ray selection. A key question is to what extent these correlations depend on galaxy parameters (e.g., type, metallicity, and environment). Once this dependence is understood we will be able to use the integrated X-ray luminosity of galaxies as an independent star-formation rate indicator. A large sample (at least thousands) of detections of normal/starburst galaxies is needed to establish these correlations over a large range of mass and star-formation in galaxy type and density subclasses. Nearby galaxies ($z < \sim 0.1$) will be detected by WFXT in large numbers, on the order of $10^{4-5}$ (see Figure 6 and Ptak et al. 2008, AIPC, 1082, 104). Distances and galactic parameters will be known from matches with surveys such as 2MASS, SDSS, Galex, Pan-Starrs and LSST. These data will also allow the X-ray logN-logS and luminosity function for for different types of normal galaxies (starbursts, spirals, elliptical galaxies) to be determined with high precision for the first time.

The resulting analysis will be on a statistical footing comparable to current SDSS-Galex studies. Note that in addition to allowing for extensive studies of galactic properties, the large SDSS-Galex database has also resulted in new classes of sources such as Ultraviolet

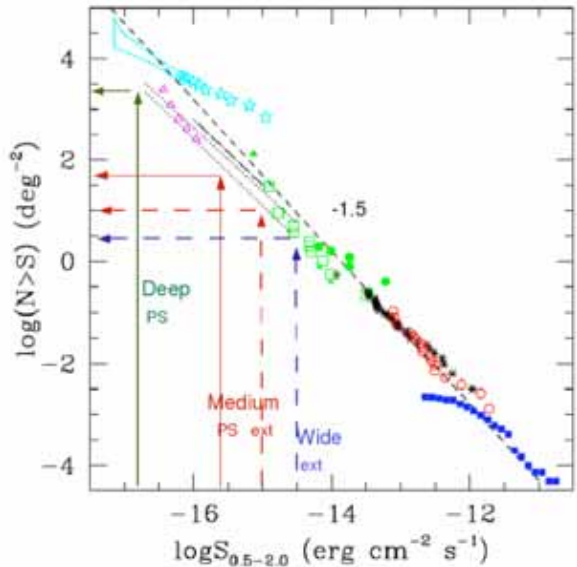

Luminous Galaxies (UVLGs; Heckman et al. 2005, ApJ, 169, 35L) which are likely local analogs to Lyman-break galaxies at z>2. *Exotic discoveries such as this are often only enabled with very wide, blind surveys.*

**Figure 6:** LogN-logS of normal/starburst galaxies (from Tajer et al. 2005, A&A, 435, 799) with the extended source (1' region) detection limits shown for the medium and wide surveys and the point-source detection limits shown for the medium ($3 \times 10^{-16}$ ergs s$^{-1}$ cm$^{-2}$) and deep ($3 \times 10^{-17}$ ergs s$^{-1}$ cm$^{-2}$) surveys.